\documentclass[conference]{IEEEtran}
\IEEEoverridecommandlockouts
\usepackage{cite}
\usepackage{amsmath,amssymb,amsfonts}
\usepackage{algorithmic}
\usepackage{graphicx}
\usepackage{textcomp}
\usepackage{xcolor}
\usepackage{multirow}
\usepackage{subcaption}
\usepackage{url}
\def\BibTeX{{\rm B\kern-.05em{\sc i\kern-.025em b}\kern-.08em
    T\kern-.1667em\lower.7ex\hbox{E}\kern-.125emX}}
\begin{document}

\title{Achieving Fairness in DareFightingICE Agents Evaluation Through a Delay Mechanism\\
}

\makeatletter
\newcommand{\linebreakand}{%
  \end{@IEEEauthorhalign}
  \hfill\mbox{}\par
  \mbox{}\hfill\begin{@IEEEauthorhalign}
}
\makeatother

\author{\IEEEauthorblockN{Chollakorn Nimpattanavong, Thai Van Nguyen, Ibrahim Khan}
\IEEEauthorblockA{\textit{Graduate School of Information Science and Engineering} \\
\textit{Ritsumeikan University, Japan} \\
\{gr0608sp, gr0557fv, gr0556vx\}@ed.ritsumei.ac.jp}
\and
\IEEEauthorblockN{Ruck Thawonmas}
\IEEEauthorblockA{\textit{College of Information Science and Engineering} \\
\textit{Ritsumeikan University, Japan} \\
ruck@is.ritsumei.ac.jp}
\linebreakand 
\IEEEauthorblockN{Worawat Choensawat, Kingkarn Sookhanaphibarn}
\IEEEauthorblockA{\textit{School of Information Technology and Innovation} \\
\textit{Bangkok University, Thailand}\\
\{worawat.c, kingkarn.s\}@bu.ac.th}
}


\IEEEoverridecommandlockouts
\IEEEpubid{\makebox[\columnwidth]{979-8-3503-2277-4/23/\$31.00~\copyright2023 IEEE \hfill} 
\hspace{\columnsep}\makebox[\columnwidth]{ }}
\maketitle
\IEEEpubidadjcol

\begin{abstract}
This paper proposes a delay mechanism to mitigate the impact of latency differences in the gRPC framework---a high-performance, open-source universal remote procedure call (RPC) framework---between different programming languages on the performance of agents in DareFightingICE, a fighting-game research platform. The study finds that gRPC latency differences between Java and Python can significantly impact real-time decision-making. Without a delay mechanism, Java-based agents outperform Python-based ones due to lower gRPC latency on the Java platform. However, with the proposed delay mechanism, both Java-based and Python-based agents exhibit similar performance, leading to a fair comparison between agents developed using different programming languages. Thus, this work underscores the crucial importance of considering gRPC latency when developing and evaluating agents in DareFightingICE, and the insights gained could potentially extend to other gRPC-based applications.
\end{abstract}

\begin{IEEEkeywords}
Fighting Game, DareFightingICE, Delay mechanism, Agent evaluation, Fair comparison
\end{IEEEkeywords}

\section{Introduction}
DareFightingICE~\cite{b1} is a Java-based game artificial intelligence (AI) research platform that aims to create a generalized fighting game agents. It builds upon the FightingICE~\cite{b2} framework and seeks to develop an agent that can learn and perform well against different opponents. One of DareFightingICE's objectives is to include the integration of audio data into the decision-making process during gameplay. By utilizing audio data, the project aims to provide an agent with the useful information that can be used to make more informed decisions. The successful integration of audio data into an agent's decision-making process could represent a significant breakthrough in game AI research.

gRPC Remote Procedure Call (gRPC)~\cite{b3} has recently been integrated into DareFightingICE~\cite{b4} as the communication framework between an agent and the platform itself. This decision was made due to the strict 16.66 millisecond time limit within which an agent must process. gRPC was chosen for its high performance in efficiently transmitting a constant stream of data. Additionally, gRPC's support for multiple programming languages opens up opportunities to develop an agent in languages other than Java. By leveraging gRPC, the project aims to improve the performance and flexibility of the agent development process. This integration is expected to provide faster and more efficient communication between an agent and the platform, ultimately leading to higher agent performance in game playing scenarios.

DareFightingICE offers an interface for developing agents in Java and Python, which are the primary languages supported by the team. However, the latency of the gRPC framework differs between the two languages, with Java exhibiting lower latency in gRPC calls to the platform compared to Python's gRPC that is not good at streaming RPC calls~\cite{b5}. This difference in latency can have a significant impact on the agent performance since the time pool available for processing on Java and Python is different. According to the game's specification, if the overall processing time exceeds 16.66 ms~\cite{b2}, an agent will not be able to process the next frame data due to frame skipping, which can lead to a drop in performance.

To ensure fairness between Java-based and Python-based agents, we propose a delay mechanism that can mitigate the effects of the latency differences between the two languages. The proposed mechanism is designed to provide a consistent processing time for both Java and Python, ultimately leading to a fair comparison between agents developed using different programming languages.

The contributions of this work are as follows: first, we conduct an investigation into gRPC latency across multiple programming languages, with a particular focus on Java and Python, and identify an appropriate delay to minimize any differences; second, we investigate the effects of latency variations on the performance of agents created for DareFightingICE by utilizing BlackMamba, the 2021 competition winner, as a test-bed.

\section{Related Work}

\subsection{Recent Studies on FightingICE and DareFightingICE Agents}
Moon et al.~\cite{b6} used a machine learning algorithm to dynamically adjust the behavior of FightingICE agents in response to players' affective states. This approach facilitated adaptive interaction between the agent and player, significantly enhancing the overall gaming experience by creating a more immersive and responsive environment. In the same year, Waris et al.~\cite{b7} utilized CATNeuro, a neural-network model based on the graph evolution concept propelled by Cultural Algorithms, to engineer real-time industrial controllers. CATNeuro was tested on FightingICE and a trailer motion system, with it consistently outperforming other methods. The superior performance of CATNeuro can be attributed to its design which fosters increased diversity in the model, a result of the interplay between cooperative and competitive knowledge.

In a separate study, Thai et al.~\cite{b8} presented a deep reinforcement learning agent for DareFightingICE. Uniquely, this agent uses sound exclusively as an input, marking a significant deviation from traditional decision-making processes used by FightingICE agents, which typically rely on game states provided by the game system, as done in the above two studies. 

\subsection{Improving Data Transfer Efficiency for Agents in the DareFightingICE using gRPC}
One of the challenges faced by agent developers in DareFightingICE as well as FightingICE was the high overhead associated with the previous communication interface used until the 2022 competition, Py4J, when transmitting large amounts of data. This often resulted in an agent's overall processing time exceeding the 16.66 ms time limit, which is crucial for timely and accurate decision-making in the competition. To address this issue, the competition has recently integrated gRPC as a communication framework between agents and DareFightingICE. The use of gRPC has several advantages, including its high performance, which reduces latency by up to 65\%~\cite{b4}, and increased stability. Moreover, gRPC has been found to mitigate the issue of missed frames, which can occur when using Py4J. The adoption of gRPC as the communication framework in DareFightingICE has significantly improved the agent performance, allowing developers to create more complex agents that can better utilize audio data in their decision-making processes.

\section{Methodology}
In this section, we elaborate on the methodology adopted to identify the optimal delay mechanism for mitigating the impact of latency differences between Java's gRPC and Python's gRPC by implementing an agent called Sandbox in both Java and Python. This involves discussing the objectives and implementation of the agent, the experimental setup to ensure accuracy and reliability, and the evaluation of latency for data transmission between the agent and the platform in both Java and Python.

\subsection{Objectives and Implementation}
Our primary objective was to pinpoint the most efficient delay mechanism for mitigating the impact of latency differences between Java's gRPC and Python's gRPC. To achieve this goal, we implemented agents both in Java and Python, named SandBox, in such a way that it would measure only the overhead on the round-trip latency of transmitting data between the agent and the platform, without processing any data. By doing so, we could isolate the delay caused exclusively by data transmission, allowing us to assess and identify the most effective delay mechanism. This approach ensures that we focus on the optimization of data transmission and communication between an agent and the platform.

\subsection{Experimental Setup}
To conduct our experiments, we employed a computer with specifications closely matching those of the official competition PC used in the DareFightingICE Competition. This similarity was crucial to ensure the accuracy and reliability of our results, given that the agent performance would be evaluated under similar conditions. The computer was equipped with an Intel(R) Xeon(R) W-2125 @ 3.70GHz CPU, 16 GB DDR4 RAM, and an NVIDIA Quadro P1000 graphics card, running on the Windows 10 Pro for Workstations operating system. Utilizing the same PC for all experiments allowed us to maintain consistent conditions and effectively eliminate other factors, which in turn facilitated an accurate comparison of the performance of the different delay mechanisms and programming languages.


\subsection{Evaluation of Latency}
\begin{figure}[b]
    \centering
    \includegraphics[width=\linewidth]{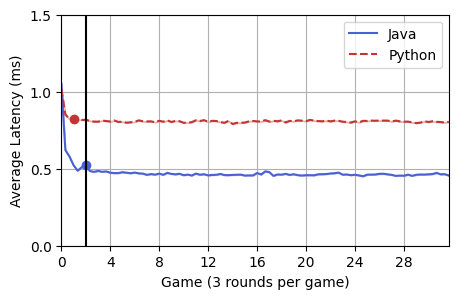}
    \caption{Sandbox's Overall Latency Comparison}
    \label{fig:sandbox}
\end{figure}

To evaluate latency, we deployed Sandbox for 32 games (96 rounds) in both Java and Python, measuring the average latency of each round and illustrating our findings in Fig.~\ref{fig:sandbox}. During the experiment, we observed that Java-based Sandbox's latency stabilized after 6 rounds, while the Python-based one's stabilized after just 3 rounds. To maintain consistency and ensure a fair comparison, we only considered the latency values after 6 rounds, at which point both agents' latency had stabilized.

The average latency for the Java-based agent after round 6 was 0.465 ms, while for Python-based Sandbox, it was 0.807 ms. We rounded the difference in latency between the two agents up to 0.35 ms. This comparison enabled us to identify the most efficient delay mechanism for Java-based agents, and ensure that this delay mechanism operates efficiently within the competition context.

\section{Evaluation}
In this section, we discuss the evaluation process designed to investigate the impact of the delay mechanism on the agent performance in the context of the DareFightingICE Competition. We outline our experimental approach, the use of BlackMamba as a test-bed, and the implementation of four variants. Furthermore, we introduce a novel evaluation method for the agent performance and present our findings, which highlight the effectiveness of the delay mechanism in mitigating performance differences between Java-based and Python-based agents.

\subsection{Experimental Approach}
Our experiments aim to investigate the impact of gRPC latency on the agent performance in DareFightingICE and the effectiveness of the delay mechanism in mitigating performance differences between Java-based and Python-based agents. To examine the impact of the delay mechanism on the agent performance, we selected BlackMamba, the winner of the 2021 FightingICE Competition, as our test-bed. We reimplemented BlackMamba in both Java and Python to enable a comparison of the performance of Java-based and Python-based agents.

The motivation behind the reimplementation is due to the fact that the initial implementation of BlackMamba in Java involved creating a new Java object in every frame, without reusing the available object. This led to frequent execution of Java Garbage Collection, resulting in unstable latency. Furthermore, since the initial implementation of BlackMamba is in Java, it is necessary to re-implement the same algorithm in Python as well. Our experiments involve four variants: the reimplemented BlackMamba (baseline) and three versions of the baseline, in both Java and Python, with processing times adjusted to 15.15 ms, 15.5 ms, and 15.85ms, respectively, as shown in Table~\ref{tab:blackmamba} in order from top to bottom.

\begin{table}[]
\caption{Implemented variants}
\label{tab:blackmamba}
\begin{center}
\begin{tabular}{|c|c|c|c|c|}
\hline
Programming & gRPC & Processing & Total (ms) \\
Language & Latency (ms) & Time (ms) & \\
\hline
\multirow{4}{3em}{Java} & \multirow{4}{2em}{0.5} & 1.1 & 1.60 \\
 &  & 15.15 & 15.65 \\
 &  & 15.50 & 16.00 \\
 &  & 15.85 & 16.35 \\
\hline
\multirow{4}{3em}{Python} & \multirow{4}{2em}{0.85} & 1.3 & 2.15 \\
 &  & 15.15 & 16.00 \\
 &  & 15.50 & 16.35 \\
 &  & 15.85 & 16.70 \\
\hline
\end{tabular}
\end{center}
\end{table}

To ensure a fair comparison, we conducted 96 rounds for each variant, where they fought against MctsAi~\cite{b9}, a sample agent in the competition using Monte-Carlo tree search. The first six rounds were disregarded to ensure consistent gRPC latency, as mentioned in Sec. III-C. Our hypothesis is that Java-based BlackMamba would outperform the Python-based one without an introduced delay in cases where the overall processing time exceeds 16.66 ms, but with a 0.35 ms delay introduced to the Java-based agents, both Java-based and Python-based agents would exhibit similar performance.

\subsection{Evaluation Method}
We introduce the method for evaluating the agent performance, taking into account both the remaining Health Points (HP) and elapsed time. The evaluation method used in existing work \cite{b1} solely focused on the remaining HP of both players, which is insufficient in effectively assessing performance. The elapsed time is also crucial as it reflects how fast an agent defeats the opponent, a factor that cannot be accurately determined by HP alone. Therefore, the elapsed time together with a boolean flag representing win or loss is also included. The reason for the introduction of this boolean flag is that if the elapsed time is the same and the fight is highly competitive with tiny HP difference, ignoring the fight result would lead to a similar assessment of performance. This new method allows for more precise and efficient evaluations.

To implement this evaluation method, the result data from each round are used, which provide information on the remaining HP for each agent and the elapsed time measured in frames, with a maximum of 3600 frames per round. These values are then normalized using a set of equations (Eqns. \eqref{eqn:HP_1}, \eqref{eqn:HP_2}, \eqref{eqn:w}, and \eqref{eqn:t}). In these equations, $HP_{BlackMamba}$ and $HP_{Mcts}$ denote the remaining HP of BlackMamba and MctsAi, respectively, while $Time_{Elapsed}$ denotes the elapsed time of the round. In addition, $HP_{Total}$ and $Time_{Total}$ denote the maximum possible HP (set at 400 HP) and the total time per round (set at 3600 frames), respectively. The average of the values from these four equations is then calculated to evaluate the performance score of BlackMamba (Eqn. \eqref{eqn:score}).

\begin{equation}
    HP_1 = \frac{HP_{BlackMamba}}{HP_{Total}}\label{eqn:HP_1}
\end{equation}

\begin{equation}
    HP_2 = 1 - \frac{HP_{MctsAi}}{HP_{Total}}\label{eqn:HP_2}
\end{equation}

\begin{equation}
    w = 
    \begin{cases}
    1, & \text{if } HP_{BlackMamba} > HP_{MctsAi}\\
    0, & \text{otherwise}
    \end{cases}
    \label{eqn:w}
\end{equation}

\begin{equation}
    t = w(1 - \frac{Time_{Elapsed}}{Time_{Total}}) + (1 - w) \frac{Time_{Elapsed}}{Time_{Total}}\label{eqn:t}
\end{equation}

\begin{equation}
    Score = \frac{HP_1 + HP_2 + w + t}{4}\label{eqn:score}
\end{equation}

\subsection{Results}
\begin{figure*}[]
    \begin{subfigure}{0.49\textwidth}
        \includegraphics[width=\textwidth]{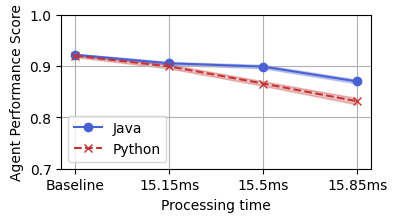}
        \subcaption{Without the Delay Mechanism}
    \end{subfigure}
    \begin{subfigure}{0.49\textwidth}
        \includegraphics[width=\textwidth]{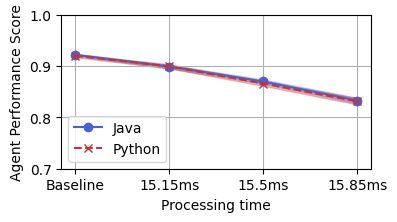}
        \subcaption{With the Delay Mechanism}
    \end{subfigure}
    \caption{Agent performance comparison between with/without the delay mechanism}
    \label{fig:blackmamba}
\end{figure*}

In Fig.~\ref{fig:blackmamba}, we observed that without the delay mechanism, Java-based BlackMamba consistently outperformed the Python version if the processing time is above 15.5 ms due to the lower gRPC latency on the Java platform. However, by adding a delay of 0.35 ms to the Java version, the performance gap was effectively reduced, with both languages showing similar performance. These results support our hypothesis that the delay mechanism can mitigate the impact of gRPC latency differences and ensure a fair and accurate evaluation of the agent performance in DareFightingICE.


\section{Discussions}
The findings indicate that both Java-based and Python-based agents demonstrate comparable performance when the processing time is under 15.15 ms, even without the introduced delay mechanism. The processing time limit that triggers the difference in the agent performance was discovered to be 15.5 ms, not 16.66 ms as mentioned in \cite{b2}. This is because our results based on average processing time, which may overlook occasional delays, resulting in the overall processing time on the game system greater than 16.66 ms.

While our experiments with BlackMamba provided valuable insights, it is important to recognize that our evaluation was limited to a single type of agent in a specific environment. Further studies are necessary to explore the effects of gRPC latency on other agents in various settings. Additionally, our study focused solely on the impact of gRPC latency differences between Java and Python, and did not consider other factors that could affect the agent performance such as operating system (OS) thread management and other OS features. Therefore, future research should investigate the impact of other variables on an agent's performance in DareFightingICE and similar applications.

\section{Conclusions}
Our study sought to investigate the impact of gRPC latency differences between programming languages on the agent performance in DareFightingICE. Specifically, we compared the performance of Java-based and Python-based agents with and without a delay mechanism. The results showed that the differences in gRPC latency between these programming languages can have a significant impact on the agent performance in DareFightingICE. However, with a delay mechanism introduced to Java-based agents, both Java-based and Python-based agents exhibited similar performance, indicating that this delay mechanism can effectively mitigate the impact of gRPC latency differences.

These findings have important implications for the development and evaluation of agents in DareFightingICE and other gRPC-based applications. When designing a game-playing AI competition that supports multiple programming languages by utilizing gRPC, it is crucial to consider the potential latency differences between programming languages and take measures to mitigate these variations. Introduction of the delay mechanisms is one such measure that can help ensure a fair and accurate evaluation of the agent performance.


\appendix

\section{Online Resources}

Source code and raw data are available at \url{https://github.com/Staciiaz/cog2023-darefightingice-evaluation}.

\end{document}